\documentclass[runningheads]{llncs}
\usepackage{url}
\usepackage[small]{caption}
\usepackage{graphicx}
\usepackage{amssymb}

\usepackage{amsthm}
\usepackage{hyperref}
\usepackage{multirow}
\usepackage{adjustbox}

\usepackage{tabu}	
\usepackage{booktabs}

\pdfoutput=1
\usepackage{amsmath}

\newtheorem{exmp}{Example}

\theoremstyle{definition}
\newtheorem{dfnt}{Definition}

\begin{document}
\title{A Quantum Annealing Algorithm for Finding Pure Nash Equilibria in Graphical Games}
\titlerunning{A Quantum Annealing Algorithm for Finding PNE-GG}
%
\author{Christoph Roch,
	Thomy Phan,
	Sebastian Feld, \\
	Robert Müller,
	Thomas Gabor,
	Carsten Hahn	   \\ and
	Claudia Linnhoff-Popien
}

\institute{
	LMU Munich\\
	\email{christoph.roch@ifi.lmu.de }}

\authorrunning{Roch et al.}
\maketitle              
\begin{abstract}
We introduce Q-Nash, a quantum annealing algorithm for
the NP-complete problem of finding pure Nash equilibria in graphical games. The algorithm consists of two phases. The first phase determines all combinations of best response strategies for each player using classical computation. The second phase finds pure Nash equilibria using a quantum annealing device by mapping the computed combinations to a quadratic unconstrained binary optimization formulation based on the Set Cover problem. We empirically evaluate Q-Nash on D-Wave's Quantum Annealer 2000Q using different graphical game topologies. The results with respect to solution quality and computing time are compared to a Brute Force algorithm and the Iterated Best Response heuristic.
\end{abstract}
\section{Introduction}

Applications of conventional game theory have played an important role in many modern strategic decision making processes including diplomacy, economics, national security, and business  \cite{carfi2011fair,rabin1993incorporating,roy2010survey}. 
Game theory is a mathematical paradigm in which such domain-specific decision situations are modeled \cite{fudenberg1991game}. Multiple players interact with each other to collectively complete a task or to enforce their interests. In the classical model of game theory \cite{von1944theory} all players choose an action simultaneously and obtain a certain payoff (\textit{utility}), which depends on the actions of the other players. The most common solution concept for such a decision problem is called \textit{Nash equilibrium (NE)} \cite{nash1951non}, in which no player is able to unilaterally improve his payoff by changing his chosen action. 

There exist many different representations for such simultaneous games. The most popular is the strategic or standard normal-form game representation, which is often used for 2-player games, like the prisoner dilemma or battle of the sexes. However, due to its exponential growth of the representational size w.r.t the number of players \cite{kearns2013graphical}, a more compact version, called graphical game, is increasingly used to model multi-player scenarios \cite{la2000game,koller2003multi,littman2002efficient,palmieri2017constraint}. Here a player's action only depends on a certain number of other players' actions (a so called player's neighborhood). These neighborhoods are visualized by an underlying graph with players as vertices and the dependencies as edges.  

While \textit{pure strategies}, where each player unambiguously decides on a particular action, are conceptually simpler than \textit{mixed strategies}, the associated computational problems appear to be harder \cite{gottlob2005pure}. This also applies to the compact representation of graphical games, for which the complexity of finding pure strategy Nash equilibria (PNE-GG) was proven to be NP-complete, even in the restricted case of neighborhoods of maximum size 3 with at most 3 actions per player \cite{gottlob2005pure}. 

Although there are many algorithms that find \textit{mixed} or \textit{approximated} NE in graphical games \cite{kearns2013graphical,ortiz2003nash,bhat2004computing,soni2007constraint}, there are only a couple of algorithms that deal with \textit{pure Nash equilibria (PNE)}. In this paper, we focus on the computationally hard problem of finding NE with \textit{pure strategies}, where each player chooses to play an action in a deterministic, non-random manner. 

With D-Wave Systems releasing the first commercially available quantum annealer in 2011\footnote{\href{https://www.dwavesys.com/news/d-wave-systems-sells-its-first-quantum-computing-system-lockheed-martin-corporation}{https://www.dwavesys.com/news/d-wave-systems-sells-its-first-quantum-computing-system-lockheed-martin-corporation}}, there is now the possibility to find solutions for such complex problems in a completely different way compared to classical computation. To use D-Wave’s quantum annealer, the problem has to be formulated as a \textit{quadratic unconstrained binary optimization (QUBO)} problem \cite{boros2007local}, which is one possible input type for the annealer. In doing so, the metaheuristic \textit{quantum annealing} seeks to find the minimum of an objective function, i.e., the best solution of the defined configuration space \cite{mcgeoch2014adiabatic}.

In this paper, we propose the first quantum annealing algorithm for finding PNE-GG, called \textit{Q-Nash}.
The algorithm consists of two phases. The first phase determines all combinations of \textit{best response} strategies for each player using classical computation. The second phase finds pure Nash Equilibria using a quantum annealing device by mapping the computed combinations to a QUBO formulation based on the Set Cover problem. We empirically evaluate Q-Nash on D-Wave's Quantum Annealer 2000Q using different graphical game topologies. The results with respect to solution quality and computing time are compared to a Brute Force algorithm and the Iterated Best Response heuristic.

\section{Background}
\subsection{Graphical Games and Pure Nash Equilibria}
In an \textit{$n$-player game}, each player $p\ (1 \leq p \leq n)$, has a finite set of strategies or actions, $S_p$, with $|S_p|\geq 2$. Such a game can be visualized by a set of $n$ matrices $M_p$. The entry $M_p(s_1,...,s_n) = M_p(s)$ specifies the payoff to player $p$ when the \textit{joint action} (also, \textit{strategy profile}) of the $n$ players is $s \in {S}$, with $S=\prod_{i=1}^{n}S_i$ being the set of combined strategy profiles. In order to specify a game with $n$ players and $s$ strategies each, the representational size is $ns^{n}$, an amount of information exponential with respect to the number of players. However, players often interact only with a limited number of other players, which allows for a much more succinct representation. In \cite{kearns2013graphical} such a compact representation, called \textit{graphical game}, is defined as follows: 

\begin{dfnt}\textbf{(Graphical Game).} An \textit{$n$-player graphical game} is a pair $(G, M)$, where $G$ is an undirected graph with $n$ vertices and $M$ is a set of $n$ matrices $M_{p}$ with $1 \leq p \leq n$, called the \textit{local game} matrices. Player $p$ is represented by a vertex labeled $p$ in $G$. We use $N_{G}(p) \subseteq \{1,...,n\}$ to denote the set of \textit{neighbors of player} $p$ in $G$ -- i.e., those vertices $q$ such that the undirected edge $(p, q)$ appears in $G$. By convention, $N_{G}(p)$ always includes $p$ himself. The interpretation is that each player is in a game with only his neighbors in $G$. Thus, the size of the graphical game representation is only exponential in the maximal node degree $d$ of the graph, $ns^{d}$. If $|N_{G}(p)|= k$ and $s \in \prod_{i=1}^{k}S_i$, $M_p(s)$ denotes the payoff to $p$ when his $k$ neighbors (including himself) play $s$.
\end{dfnt} 

Consider a game with $n$ players and strategy sets $S_1,...,S_n$. For every strategy profile $s\in S$, the strategy of player $p$ is denoted by $s_p$ and $s_{-p}$ corresponds to the $(n-1)$-tuple of strategies of all players but $p$. For every $s_{p}^{'} \in S_p$ and $s_{-p} \in S_{-p}$ we denote by $(s_{-p};s_{p}^{'})$ the strategy profile in which player $p$ plays $s_{p}^{'}$ and all the other players play according to $s_{-p}$.
One has to mention that a strategy profile $s$ is called \textit{global}, if all $n$ players contribute to it, i.e., a global combined strategy $s$ consists of every player playing one of his actions \cite{gottlob2005pure}. 

\begin{dfnt}\textbf{(Pure Nash Equilibrium).} A global strategy profile $s$ is a PNE, if for every player $p$ and strategy $s_{p}^{'} \in S_p$ we have $M_{p}(s) \geq M_{p}(s_{-p};s_{p}^{'})$. That is, no player can improve his expected payoff by deviating unilaterally from a Nash equilibrium. 
\end{dfnt} 
\cite{daskalakis2006computing} define a \textit{best re\-sponse} strategy as follows: \smallskip

\begin{dfnt}\label{def:best-response}\textbf{(Best Response Strategy).} A best response strategy of player p is defined by:
	\begin{equation*}
	BR_{M_p}(s_{-p})\triangleq\{ s_p\mid s_p \in S_p \text{ and } \forall s_{p}^{'} \in S_p : M_p(s_{-p};s_p) \\ \geq M_p(s_{-p};s_{p}^{'}) \}
	\end{equation*}
	Intuitively, $BR_{M_p}(s_{-p})$ is the set of strategies in $S_p$ that maximize $p$'s payoff if the other players play according to $s_{-p}$.
	Thus, a strategy profile $s$ is a pure Nash equilibrium if for every player $p$, $s_p \in BR_{M_p}(s_{-p})$.
\end{dfnt} 

A visual example of a graphical game called \textit{FRIENDS} and its PNE can be found in \cite{gottlob2005pure}.

\subsection{Quantum Annealing}\label{sec:qauntum-annealing}

\textit{Quantum annealing} is a metaheuristic for solving complex optimization and decision problems \cite{kadowaki1998quantum}. D-Wave’s quantum annealing heuristic is implemented in hardware, designed to find the lowest energy state of a spin glass system, described by an Ising Hamiltonian,
\begin{equation}\label{eq:ising}
\mathcal{H}(s) = \sum_{i} h_{i}x_{i} + \sum_{i<j} J_{ij}x_{i}x_{j}
\end{equation}
where $h_{i}$ is the on-site energy of qubit $i$, $J_{ij}$ are the interaction energies of two qubits $i$ and $j$, and $x_{i}$ represents the spin $(-1, +1)$ of the $i$th qubit. The basic process of quantum annealing is to physically interpolate between an initial Hamiltonian $H_{I}$ with an easy to prepare minimal energy configuration (or ground state), and a problem Hamiltonian $H_{P}$, whose minimal energy configuration is sought that corresponds to the best solution of the defined problem  (see Eq. \ref{eq:process}). This transition is described by an adiabatic evolution path which is mathematically represented as function $s(t)$ and decreases from 1 to 0 \cite{mcgeoch2014adiabatic}. 
\begin{equation}\label{eq:process}	
H(t) = s(t)H_I + (1-s(t))H_P
\end{equation}
If this transition is executed sufficiently slow, the probability to find the ground state of the problem Hamiltonian is close to 1 \cite{albash2018adiabatic}. Thus, by mapping the Nash equilibrium decision problem onto a spin glass system, quantum annealing is able to find the solution of it.

For completeness, we map our NE decision problem to an alternative formulation of the Ising spin glass system. The so called QUBO problem \cite{boros2007local} is mathematically equivalent and uses 0 and 1 for the spin variables \cite{su2016quantum}. The quantum annealer is as well designed to minimize the functional form of the QUBO: 
\begin{equation}\label{eq:QUBO}
\min x^tQx \qquad \text{with } x \in \{0,1\}^n 
\end{equation}
with $x$ being a vector of binary variables of size $n$, and $Q$ being an $n \times n$ real-valued matrix describing the relationship between the variables. Given the matrix $Q:n \times n$, the annealing process tries to find binary variable assignments $ x \in \{0,1\}^n$ to minimize the objective function in Eq. \ref{eq:QUBO}.

\subsection{Set Cover Problem}\label{sec:set-cover}
Since the QUBO formulation of Q-Nash resembles the well known Set Cover (SC) problem, it is introduced here. Within the SC problem, one has to find the smallest possible number of subsets from a given collection of subsets $V_{k} \subseteq U$ with $1 \leq k \leq N$, such that the union of them is equal to a global superset $U$ of size $n$. This problem was proven to be NP-hard \cite{karp1972reducibility}. In \cite{lucas2014ising} the QUBO formulation for the Set Cover problem is given by: 

\begin{equation}\label{eq:QUBO SC-hard}
H_A = A \sum_{\alpha=1}^{n}  \left(1- \sum_{m=1}^{N} x_{\alpha,m} \right)^2 + A \sum_{\alpha=1}^{n}  \left(\sum_{m=1}^{N}mx_{\alpha,m} -\sum_{k:\alpha \in V_k} x_{k} \right)^2
\end{equation}

and

\begin{equation}\label{eq:QUBO SC-hard-2}
H_B = B \sum_{k=1}^{N} x_k
\end{equation}

with $x_k$ being a binary variable which is $1$, if set $k$ is included within the chosen sets, and $0$ otherwise.
$x_{\alpha,m}$ denotes a binary variable which is $1$ if the number of chosen subsets $V_{k}$ which include element $\alpha$
is $m \geq 1$, and $0$ otherwise. The first energy term imposes the constraints that for any given $\alpha$ exactly one
$x_{\alpha,m}$ must be $1$, since each element of $U$ must be included a fixed number of times. The second term states, that the number
of times that we claimed $\alpha$ was included is in fact equal to the number of subsets $V_k$ we have included with $\alpha$ as
an element. $A$ is a penalty value, which is added on top of the solution energy, described by $H = H_A + H_B$, if a constraint was not satisfied, i.e. one of the two terms (quadratic differences) are unequal to $0$. Therefore adding a penalty value states a solution as invalid. Additionally, the SC problem minimizes over the number of chosen subsets $V_{k}$, as stated in Eq. \ref{eq:QUBO SC-hard-2}. We skip discussing the term due to the fact that it has no impact on our Q-Nash QUBO problem later on.

\section{Related Work}

Most known algorithms focus on finding \textit{mixed} or \textit{approximated} NE \cite{kearns2013graphical,ortiz2003nash,bhat2004computing,soni2007constraint}. However, some investigations in determining PNE in graphical and similar variations of games were made.

Daskalakis and Papadimitriou present a reduction from GG to Markov random fields such that PNE can be found by statistical inference. They use known statistical inference algorithms like belief propagation, junction tree algorithm  and Markov Chain Monte Carlo to determine PNE in graphical games \cite{daskalakis2006computing}. 

In \cite{jiang2007computing}, the authors analyze the problem of computing pure Nash equilibria in action graph games (AGGs),  another compact game theoretic representation, which is similar to graphical games. They propose a dynamic-programming approach that constructs equilibria of the game from equilibria of restricted games played on subgraphs of the action graph. In particular, under the premise that the game is symmetric and the action graph has bounded treewidth, their algorithm determines the existence of a PNE in polynomial time. 

Palmieri and Lallouet deal with constraint games, for which constraint programming is used to express players preferences. They rethink their solving technique in terms of constraint propagation by considering players preferences as global constraints. Their approach is able to find all pure Nash equilibria for some problems with 200 players and also shows that performance can be improved for graphical games \cite{palmieri2017constraint}.

With quantum computing gaining more and more attention\footnote{\href{https://www.gartner.com/smarterwithgartner/the-cios-guide-to-quantum-computing/}{https://www.gartner.com/smarterwithgartner/the-cios-guide-to-quantum-computing/}} and none of the related work making use of quantum annealing in order to find PNE, we propose a solution approach using a quantum annealer. 

Since some NE algorithms are restricted to a certain graphical game structure, see for example \cite{littman2002efficient}, we want to emphasize, that our approach is able to work on every graphical game structure, even if the dependency graph is not connected.

\section{Q-Nash}
In the following sections we present the concept of Q-Nash. Q-Nash consists of two phases, which are described below.
\subsection{Determining best response strategies}\label{sec:determining-br}
In the first phase, we identify each player’s \textit{best response} to what the other players might do. That is, for every strategy profile $s$, we search player $p$`s strategy (or strategies) with the maximum payoff $M_p(s)$. This involves iterating through each player in turn and determining their optimal strategies. An example is given in \ref{exp:matrix}. This can be feasibly done in polynomial time, since one can easily explore each player's matrix $M_p$ representing the utility function, i.e., payoffs \cite{gottlob2005pure}. Therefore the first part, which we see as preliminary step for our Q-Nash algorithm, is executed on a classical computer. After doing this, one gets a set of combined strategies with each being a \textit{best response} to the other players' played strategies. This set is denoted by $\mathcal{B}=\{BR_{M_p} \mid (1 \leq p \leq n) \}$ and has the cardinality $C_\mathcal{B} = \sum_{p=1}^{n} |BR_{M_p}| $.

\begin{exmp}\label{exp:matrix}
	{\em In Tab. \ref{tab:Local_Matrix_Player_A} the local payoff matrix $M$ of player $A$ is visualized. For instance, assume player $B$ plays action $2$ and player $C$ chooses action $1$. In this case, a \textit{best response} strategy for player $A$ is action $0$, due to the fact, that he gets the most payoff in this situation, i.e. 4. This leads to a \textit{best response} strategy combination for player $A$, denoted by a pointed set $\{A0, B2, C1\}$ with player $A$ being the base point of it.}
\end{exmp}

\setlength{\tabcolsep}{5pt}
\begin{table}[h]
	\centering

		\begin{tabular}{c|cccccc}
			\toprule
			A & B0 C0 & B0 C1 & B1 C0 & B1 C1 & B2 C0 & B2 C1 \\
			\midrule
			0 & \textbf{4} & 1 & \textbf{2} & \textbf{2} & 1 & \textbf{4} \\
			1 & 1 & \textbf{3} & \textbf{2} & 1 & \textbf{2} & 2 \\
			\bottomrule
	\end{tabular}
	\vspace{+1em}
	\caption{Local dependency payoff matrix $M$ of player $A$. The payoffs marked in bold correspond to a \textit{best response} strategy of player $A$.}
	\label{tab:Local_Matrix_Player_A}
\end{table}
\subsection{Finding PNE using quantum annealing}
In the second phase, a PNE is identified, when all players are playing one of their \textit{best response} strategies simultaneously. 
With the computed set $\mathcal{B}$ of the classical phase, the following question arises:

\vspace{+1em}
\noindent\textit{``Is there a union of the combined \textit{best response} strategies of $\mathcal{B}$ which results in a global strategy profile, under the premise that every player plays one of his best response actions?''} --- As stated in Definition \ref{def:best-response} this would lead to a PNE. 

\vspace{+1em}
This question resembles the Set Cover (SC) problem, stated in Sec.~\ref{sec:set-cover}. It asks for the smallest possible number of subsets to cover the elements of a given global set (in our case, this global set would be a feasible global strategy profile and the subsets correspond to our best response strategy profiles of $\mathcal{B}$). However, for our purposes we have to modify the given formulation in Eq. \ref{eq:QUBO SC-hard} as follows:
\begin{equation}\label{eq:QUBO SC-complete}
\begin{gathered}
H = A \sum_{p=1}^{n}  \left(1- \sum_{j=1}^{|S_p|} \sum_{m=1}^{C_\mathcal{B}} x_{p,j,m} \right)^2 + A \sum_{p=1}^{n} \sum_{j=1}^{|S_p|} \left(\sum_{m=1}^{C_\mathcal{B}}mx_{p,j,m} -\sum_{k:p,j \in s_k} x_{k} \right)^2 \\ + A \left(n - \sum_{k=1}^{C_\mathcal{B}} x_k \right)^2
\end{gathered}
\end{equation}

Eq. \ref{eq:QUBO SC-complete} is quite similar to Eq. \ref{eq:QUBO SC-hard}. However, an element $\alpha$ of the superset of Eq. \ref{eq:QUBO SC-hard} corresponds to a player $p$ and his chosen action $j$ of our global strategy profile. Nevertheless, the intention of those first two energy terms complies with the intention of Eq. \ref{eq:QUBO SC-hard}, stated in Sec. \ref{sec:set-cover}. Further, another energy term must be added to our QUBO problem as constraint. This last energy term, for which an instance is given in Example \ref{exp:constraint}, states that exactly $n$ sets of $\mathcal{B}$ should be included to form the global strategy profile. This constraint implicitly ensures that every player is playing one of his best response strategies.

$A$ is called penalty value, which is added on top of the solution energy, if a constraint was not satisfied, i.e. one of the three terms (quadratic differences) are unequal to $0$. Thus, adding a penalty value states a solution as invalid. Only if the total energy described by $H=0$ the corresponding solution is a valid \textit{global best response} strategy profile and thus a PNE.

All these energy terms are specified within the QUBO problem matrix $Q$, which the quantum annealer takes as an input, goes through the annealing process and responds with a binary vector $x$ as a solution, see Sec. \ref{sec:qauntum-annealing}. This vector indicates which best response strategy of each player should be chosen to form a PNE.

\begin{figure}[h]
	\centering
	\includegraphics[width=0.6\textwidth]{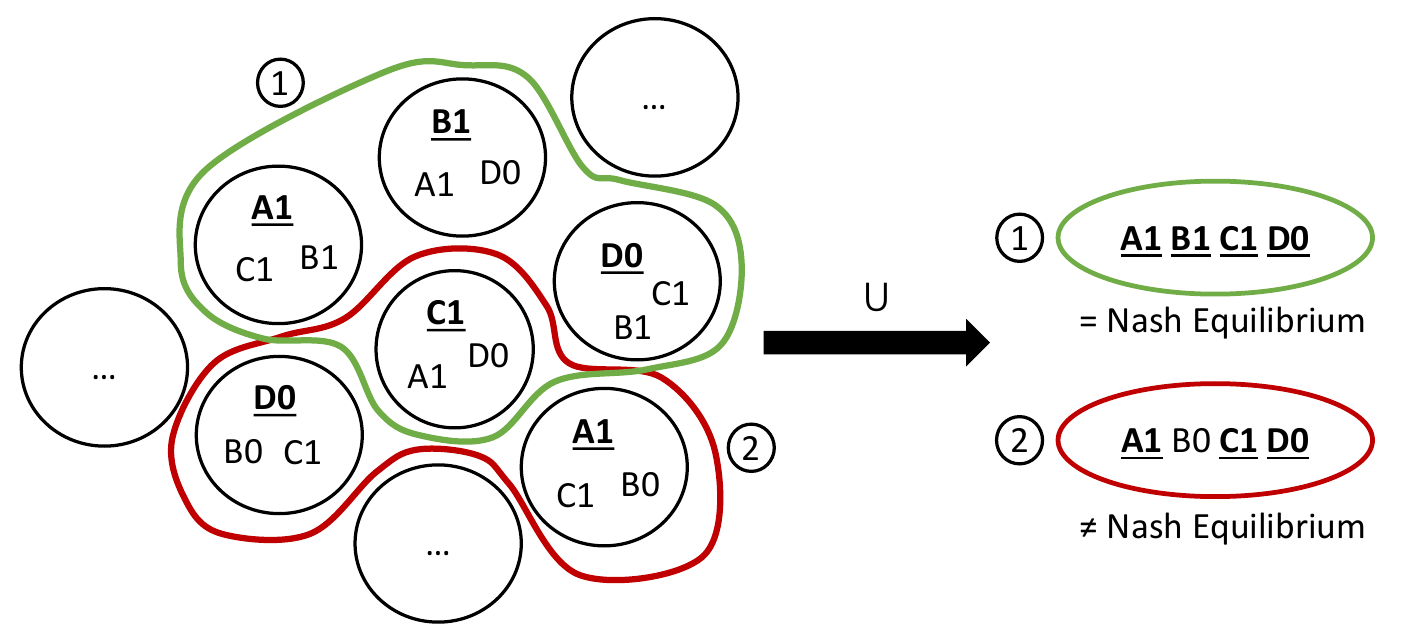}
	\caption{On the left hand an excerpt of superset $\mathcal{B}$ is given. On the right hand the sets are united, with each (1) \& (2) being a global strategy combination, but only one (1) being a PNE.}
	\label{fig:union-example}
\end{figure}

\begin{exmp}\label{exp:constraint}
	{\em To demonstrate the function of the last energy term (constraint) of Eq. \ref{eq:QUBO SC-complete}, an excerpt of \textit{best response} strategy combinations of $\mathcal{B}$ (in form of pointed sets, with the bold player-action-combination being the base point) for an arbitrary 4-player game are visualized in Fig. \ref{fig:union-example}. The green union of four \textit{best response} sets leads to a PNE, in which every player is playing one of his \textit{best response} strategies. Although the red union of three sets also leads to a global combined strategy set, in which every player is playing one of his actions, it is not a PNE, due to player $B$ not playing a \textit{best response} strategy.} 
\end{exmp}

\section{Experiments}
\subsection{Evaluation Graph-Topologies}

\begin{figure}[h]
	\centering
	\includegraphics[width=0.8\textwidth]{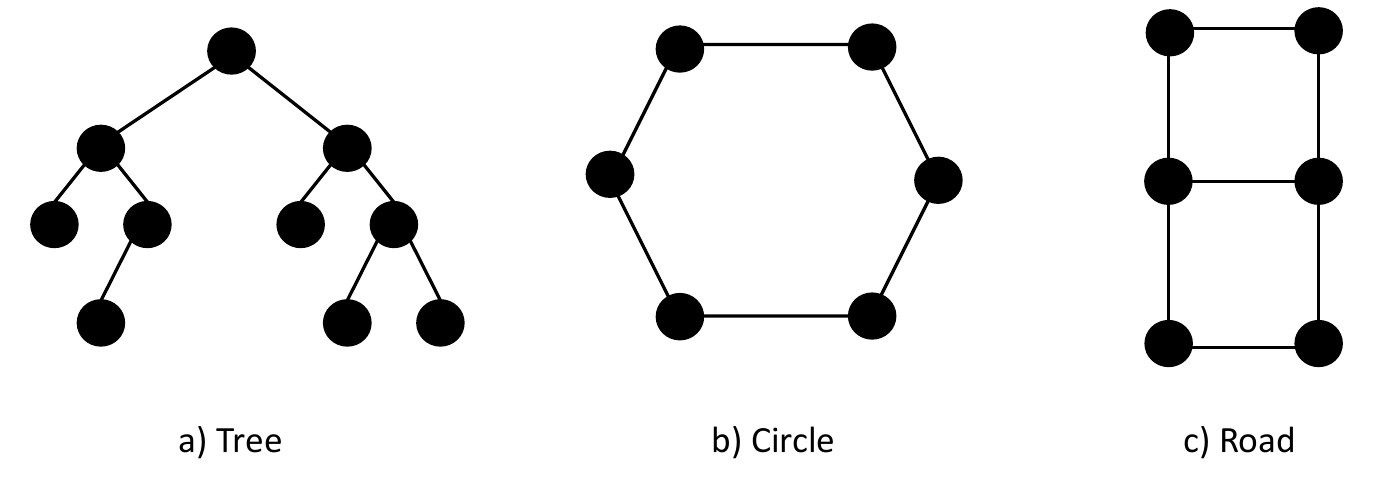}
	\caption{Different graphical game structures (topologies) which indicate the dependencies between players of a game. (a) Tree-Topology with $n=10$, (b) Circle-Topology with $n=6$ and (c) Road-Topology with $n=6$.}
	\label{fig:topologies}
\end{figure}

For evaluating Q-Nash, we implemented a game generator. It creates graphical game instances of three different popular graphical structures, which were often used in literature \cite{koller2003multi,vickrey2002multi,jiang2007computing,palmieri2017constraint}. These graphical structures are shown in Fig. \ref{fig:topologies}. As an input for our game generator, one can choose the number of players, the graph-topology and thus the dependencies between the players and the number of actions for each player individually. The corresponding payoffs are sampled randomly from [0, 15]. For our experiments we considered games with three actions per player. The classic theorem of Nash \cite{nash1951non} states that for any game, there exists a Nash equilibrium in the space of joint \textit{mixed strategies}. However, in this work we only consider \textit{pure strategies} and therefore there might be (graphical) games without any PNE (see, for instance, \cite{osborne1994course}). Additionally we want to emphasize, that Q-Nash is able to work on every graphical game structure, even if the dependency graph is not connected, for instance a set of trees (called \textit{forest}). 
\subsection{Methods}\label{sec:methods}

\subsubsection{QBSolv:}

Due to the fact, that quantum computing is still in its infancy, and corresponding hardware is limited in the number of qubits and their connectivity, we need to fall back to a hybrid method (QBSolv\footnote{\href{https://github.com/dwavesystems/qbsolv}{https://github.com/dwavesystems/qbsolv}}), in order to solve large problem instances. QBSolv is a software that automatically splits instances up into subproblems submitted to D-Wave's quantum annealer, and an extensive tabu search is applied to post-process all D-Wave solutions. Additionally, QBSolv embeds the QUBO problem to the quantum annealing hardware chip. QBSolv further allows to specify certain parameters such as the number of individual solution attempts (\textit{num\_repeats}), the subproblem size used to split up instances which do not fit completely onto the D-Wave hardware and many more. For detailed information, see \cite{booth2017partitioning}.

\subsubsection{Brute Force:}

For evaluating the effectiveness of Q-Nash we implemented a Brute Force (BF) algorithm to compare with. It determines the best response strategy sets of all players in the same way as Q-Nash does and afterwards tries out every possible combination of those sets to form a valid global strategy set which corresponds to a PNE. The number of combinations is exponential with respect to the number of players of the game, $\prod_{p=1}^n BR_{M_p}$.

\subsubsection{Iterated Best Response:}

Additionally, an Iterated Best Response (IBR) algorithm was implemented \cite{roughgarden2016twenty}. In each iteration, one player changes his action to the best action that is the best response to the other players their action. This is repeated until a PNE is reached. In that case, the algorithm starts again from an randomly generated global strategy profile, to find other PNE. Furthermore it takes a timespan as an input parameter and terminates after a timeout occured.

\subsection{Computational times of Q-Nash}\label{sec:comp-times}
\begin{figure}[h]
	\centering
	\includegraphics[width=0.7\textwidth]{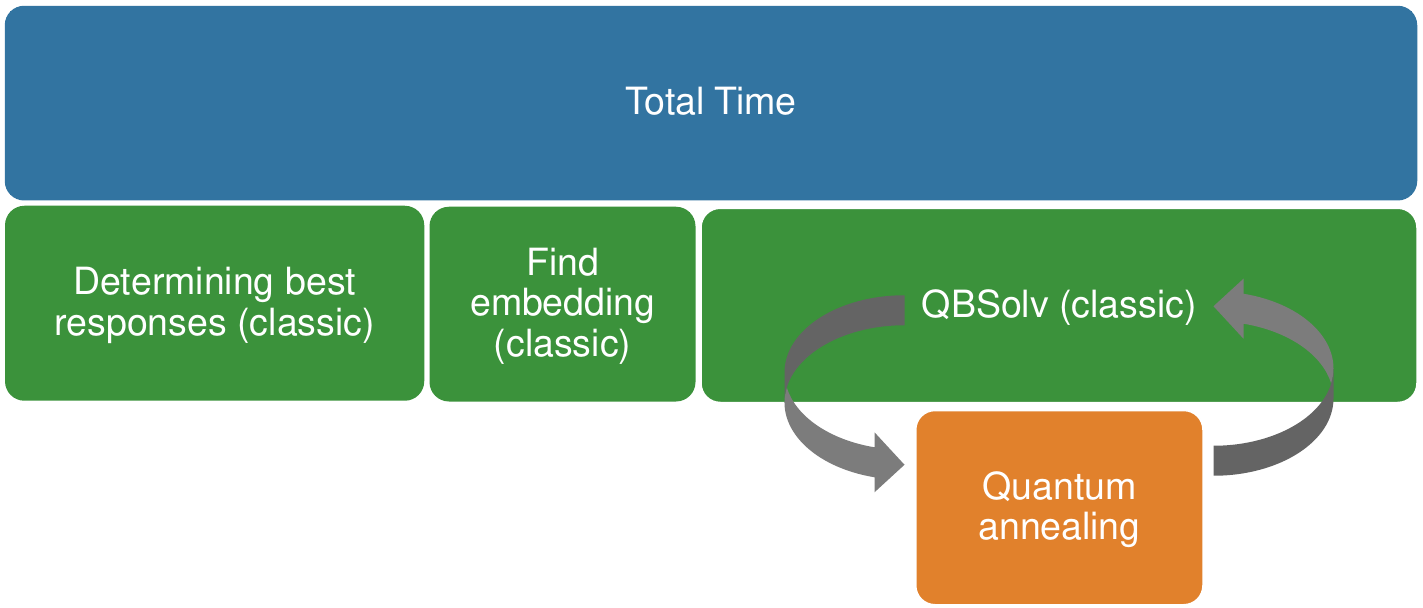}
	\caption{Overview of the computational time components of Q-Nash}
	\label{fig:time-components}	
\end{figure}

With respect to the computational results stated in Tab. \ref{tab:computational-results} we first introduce the time components of Q-Nash's total computation time. For a better understanding a general overview of Q-Nash is given in Fig. \ref{fig:time-components}.\smallskip

\textit{Determining best responses} describes the time Q-Nash classical computing phase takes to identify the \textit{best response} strategies of every player and additionally build up the QUBO matrix. The \textit{Find embedding time} states the time D-Wave's classical embedding heuristic takes to find a valid subproblem hardware embedding. The \textit{QBSolv time} can be divided into the \textit{classical time} and the \textit{quantum annealing time}. The \textit{classical QBSolv time} comprises of not only a tabu search, which iteratively processes all subproblem solutions, it also contains the latency and job queuing time to D-Wave's quantum hardware. The \textit{quantum annealing time} comprises of the number of subproblems times D-Wave's \textit{qpu-access-time}.  

\section{Results \& Discussion}\label{sec:results}

\begin{figure}[t!]
	\centering
	\begin{minipage}[c]{0.48\textwidth}
		\includegraphics[width=\textwidth]{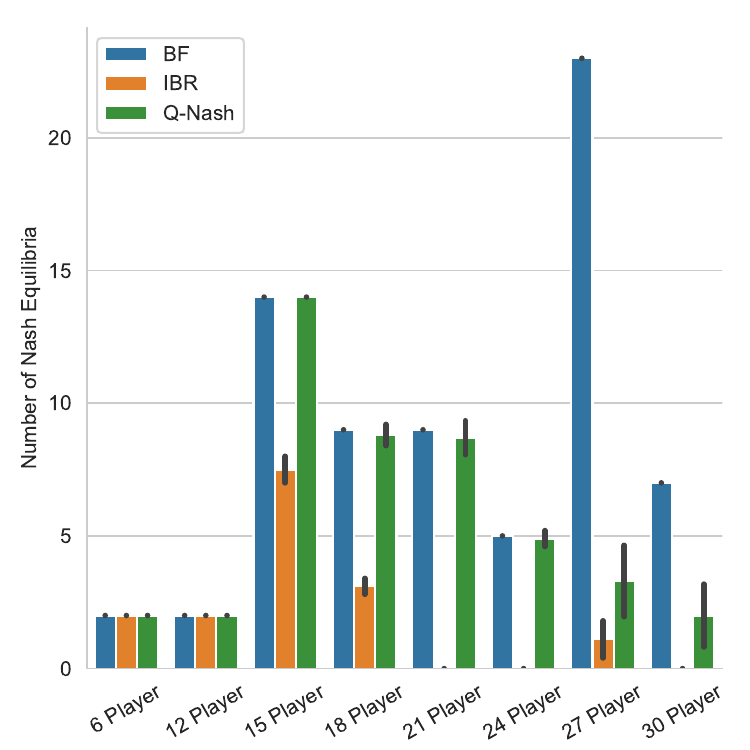}
		\caption{Circle-Topology games}
		\label{fig:circle}
	\end{minipage}
	\begin{minipage}[c]{0.48\textwidth}
		\includegraphics[width=\textwidth]{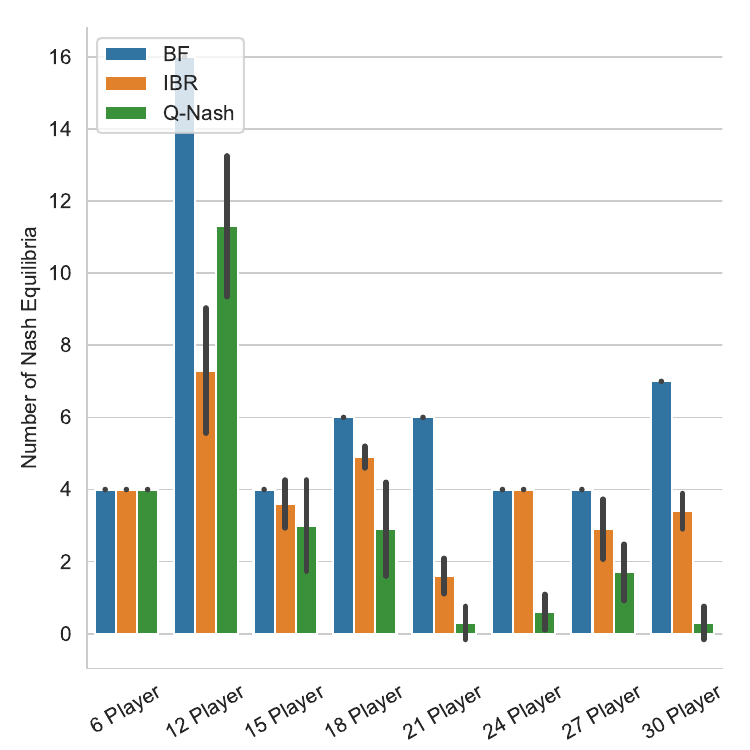}
		\caption{Tree-Topology games}
		\label{fig:tree}
	\end{minipage}
	
	\begin{minipage}[c]{0.48\textwidth}
		\includegraphics[width=\textwidth]{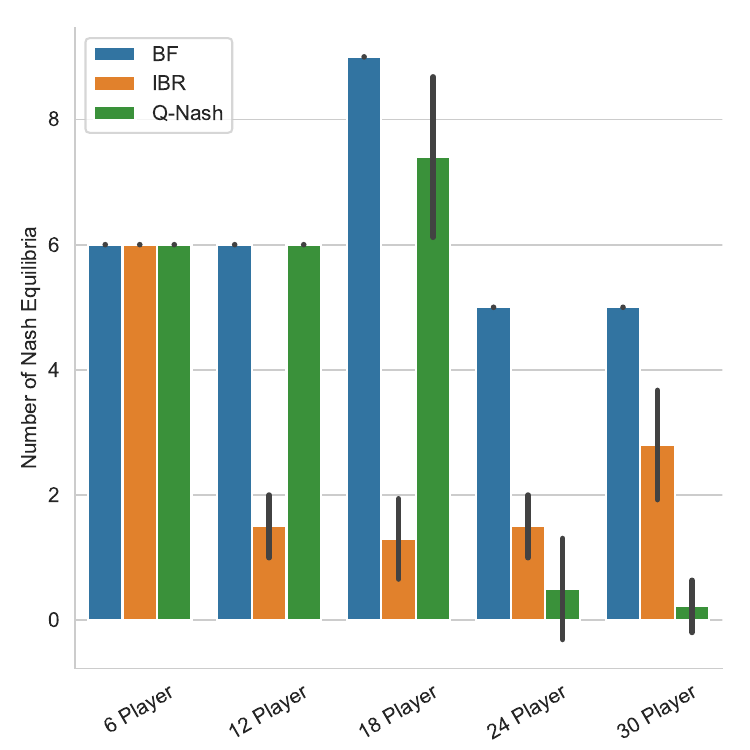}
		\caption{Road-Topology games}
		\label{fig:road}
	\end{minipage}
	\begin{minipage}[c]{0.48\textwidth}
		\includegraphics[width=\textwidth]{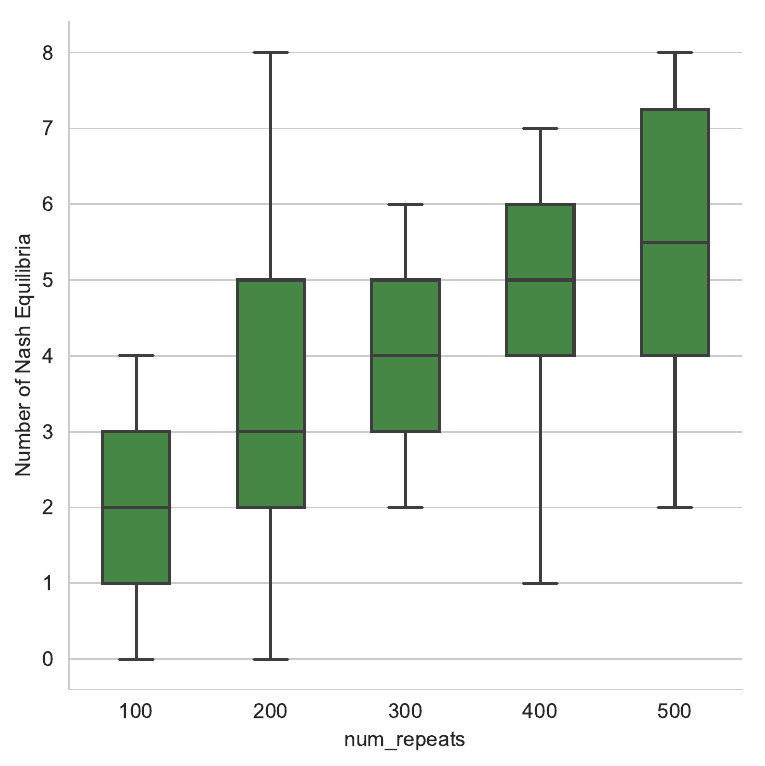}
		\caption{The impact of \textit{num\_repeats} parameter on a 24 player road game.}
		\label{fig:scaling-num-repeats}
	\end{minipage}
\end{figure}

We investigated the solution quality and computational time of Q-Nash.
Fig. \ref{fig:circle}, \ref{fig:tree} and \ref{fig:road}  show the ability of finding PNE of the three proposed methods (Q-Nash with QBSolv, BF and IBR) in differently structured graphical games. For every graph topology (Tree, Circle and Road) we used games with players ranging from 6 to 30, due to the fact that the road topology needs an even number of players we skipped 15, 21 and 27 player instances. We ran Q-Nash and IBR 20 times on every instance. Additionally, IBR was run as long as Q-Nash (total time) took, to solve the instances. 

The results show that Q-Nash with $num\_repeats$ set to 200, always found the same amount of PNE as the exact BF algorithm for the smaller game instances (6 to 12 Players, except of the 12 Player Tree-Topolgy instance), while IBR was only able to find all PNE in the Circle-Topology for those instances. 

Regarding the larger game instances, one can see that Q-Nash, due being a heuristic, was not always able to find the same amount of PNE per run (20 times) and also was not able to find every PNE in a game, when compared to the exact BF algorithm. Compared to IBR, one can notice, that Q-Nash in general performed better than IBR on Circle-Topology, while IBR outperformed Q-Nash on tree structured games. Regarding the Road-Topology, Q-Nash did better on 12 and 18 Player instances, while IBR surpassed Q-Nash on 24 and 30 Player instances.

\raggedbottom

As already mentioned in Sec. \ref{sec:methods} QBSolv splits the QUBO into smaller components (subQUBOS) of a predefined subproblem size, which are then solved independently of each other. This process is executed iteratively as long as there is an improvement and it can be defined using the QBSolv parameter \textit{num\_repeats}. This parameter determines the number of times to repeat the splitting of the QUBO problem matrix after finding a better sample. In Fig. \ref{fig:scaling-num-repeats} the influence of this parameter is shown. We exemplary used a 24 player road game and ran Q-Nash 20 times per parameter setting to show its impact on the effectiveness. As expected one can see, that with increasing number of repeats (\textit{num\_repeats}) the inter-quartile range and its median in regard to the number of PNE found, increase. Although an annealing process takes only 20$\mu s$ in default, it adds up with the \textit{num\_repeats} parameter and therefore leads to a trade-off between computational time and accuracy. 
\begin{table*}[h]
	\centering
	\begin{adjustbox}{width=\textwidth}
		\begin{tabu}{cc|[2pt]c|c|c|c|c|c|c|c|}
			\cline{3-10}
			& & \multicolumn{8}{ c|}{\rule{0pt}{10pt} \textbf{Graphical Games (Circle)}}   \\ \cline{3-10}
			& & 6 Player & 8 Player & 10 Player & 12 Player & 14 Player & 16 Player & 18 Player & 20 Player \rule{0pt}{10pt}  \\ \tabucline[2pt]{-}
			\multicolumn{1}{ |c  }{\multirow{5}{*}{\textbf{Q-Nash}} } &
			\multicolumn{1}{ |l|[2pt] }{(1) Total time [s]} & \textbf{85.393} & \textbf{125.828} & \textbf{457.906} & \textbf{424.701} & \textbf{273.283} & \textbf{228.609} & \textbf{276.593} & \textbf{248.547}  \rule{0pt}{10pt}  \\ \cline{2-10}
			\multicolumn{1}{ |c  }{} &
			\multicolumn{1}{ |l|[2pt] }{(2) Determining best responses (Classic) [s]} & 0.391 & 0.891 & 1.828 & 3.063 & 4.906 & 7.328 & 10.704 & 14.360 \rule{0pt}{10pt}   \\ \cline{2-10}
			\multicolumn{1}{ |c  }{} &
			\multicolumn{1}{ |l|[2pt] }{(3) Find embedding (Classic) [s]} & 4.172 & 4.062 & 4.093 & 5.155 & 4.813 & 5.703 & 4.202 & 4.391  \rule{0pt}{10pt}  \\ \cline{2-10}
			\multicolumn{1}{ |c  }{} &
			\multicolumn{1}{ |l|[2pt] }{(4) QBSolv (Classic) [s]} & 79.862 & 120.012 & 448.834 & 412.869 & 260.285 & 212.849 & 258.395 & 226.775 \rule{0pt}{10pt}  \\ \cline{2-10}
			\multicolumn{1}{ |c  }{} &
			\multicolumn{1}{ |l|[2pt] }{(5) QA time (Quantum) [s]} & 0.968 & 0.863 & 3.151 & 3.614 & 3.279 & 2.729 & 3.292 & 3.021  \rule{0pt}{10pt}  \\ \cline{1-10}
		\end{tabu}	
	\end{adjustbox}
	\vspace{+1em}
	\caption{Computational times of the Q-Nash algorithm on various circle structured graphical games}
	\label{tab:computational-results}
\end{table*}

The computational time results are shown for circle structured graphical games and can be viewed in Tab. \ref{tab:computational-results}. The used game instances differ in the number of players (6-20 Player) and the computing times are given in seconds of our Q-Nash algorithm. As already mentioned in Sec. \ref{sec:methods} we need to use QBSolv to solve large problem instances and therefore we have to consider the computational results of Q-Nash with caution, as stated in Sec. \ref{sec:comp-times}.


Due to Q-Nash's immense overhead time (embedding, tabu search, latency, queuing time, etc.) the \textit{total time (1)} is even for the smaller game instances (6 and 12 Players) quite large (85.393 and 125.828 sec). Nevertheless the pure Q-Nash computational time (\textit{determining best response strategies (2) \& QA time (5)}) is comparatively very small (1.359 and 1.754 sec).

In general, one can see that the \textit{quantum annealing time (5)} for these game instances is quite constant and the runtime of the \textit{classical Q-Nash phase (4)} is only polynomial, as stated in Section \ref{sec:determining-br}.  

With the \textit{embedding time (3)} being constant, the only unpredictable runtime component of Q-Nash is \textit{QBSolv classic (4)}. This is due to the tabu search (including the \textit{num\_repeats} parameter), which only terminates after a specified number of iterations in which no improvement of the previously found solutions could be made.

However, with increasing number of qubits and their connectivity it might be possible to map larger game instances directly to the hardware chip such that the hybrid solver QBSolv can be avoided. Thus, the total Q-Nash time would only consist of the \textit{classical algorithm phase time (2)}, the \textit{embedding time (3)} and a fraction of the \textit{quantum annealing time (5)}, since we do not have to split our problem instance into subproblems anymore. At this time, Q-Nash might be a competitive method regarding the computational time.

\section{Conclusion}

We proposed Q-Nash, to our knowledge the first algorithm that finds PNE-GG using quantum annealing hardware. 
Regarding the effectiveness of Q-Nash, we showed that for small game instances (ranging from 6-12 players) the algorithm was always able to find all PNE in differently structured graphical games. Anyway we have to mention, that with increasing number of players the variance w.r.t the number of found PNE increased, too.  

Due to the fact that quantum computing is still in its infancy and recent hardware is limited in the number of qubits and their connectivity, we had to fall back on a quantum-classical hybrid solver, called QBSolv, which involves additional overhead time. That makes it difficult to draw a fair comparison of Q-Nash and classical state-of-the-art solution methods regarding the computational time. We therefore decomposed the total time into its main components to show their impact. According to the experimental results, the only unpredictable time component is QBSolv's classical tabu search along with latency and job queuing times at D-Wave's cloud computing frontend. However with D-Wave announcing an immense rise of the number of qubits and their connectivity on D-Wave's quantum processors in the next years\footnote{\href{https://www.dwavesys.com/sites/default/files/mwj_dwave_qubits2018.pdf}{https://www.dwavesys.com/sites/default/files/mwj\_dwave\_\\qubits2018.pdf}}, it might be possible to embed larger game instances directly onto the chip and therefore omit hybrid solvers like QBSolv. Another possibility is using Fujitsu's Digital Annealing Unit (DAU) which also takes a QUBO matrix as input. With DAU being able to solve larger fully connected QUBO problems \cite{aramon2019physics}, a shorter total computation time could be achieved.

Regarding future work, it would also be interesting to see, how Q-Nash performs on a gate model quantum computer. For example, one could use the quantum approximate optimization algorithm, proposed by Farhi et al. \cite{farhi2014quantum}. This algorithm takes a QUBO Hamiltonian as an input. However since gate model quantum computers are at the moment even more limited in their resources, one has to come up with a clever way to split up large problem instances to fit them on the quantum chip.

%
%
%
\bibliographystyle{splncs04}
\bibliography{bibfile}

\end{document}